\documentclass[final,3p,times]{elsarticle}

\usepackage{amssymb}
\usepackage{amsmath}
\usepackage{float}

\journal{Solid State Communications}

\begin{document}

\begin{frontmatter}



\title{Study of gold and bismuth electrical contacts to a MoS$_{2}$ monolayer}


\author[first]{Agata Zielińska}
\author[second]{Joanna Prażmowska-Czajka}
\author[first]{Mateusz Dyksik}
\author[third]{Jonathan Eroms}
\author[third]{Dieter Weiss}
\author[second]{Regina Paszkiewicz}
\author[first]{Mariusz Ciorga}
\affiliation[first]{organization={Department of Experimental Physics, Wrocław University of Science and Technology},
            addressline={Wybrzeże Wyspiańskiego 27}, 
            postcode={50-370},
            city={Wrocław}, 
            country={Poland}}
            
\affiliation[second]{organization={Faculty of Electronics, Photonics and Microsystems, Wrocław University of Science and Technology},
            addressline={Janiszewskiego 11/17}, 
            postcode={50-372},
            city={Wrocław}, 
            country={Poland}}

\affiliation[third]{organization={Institute for Experimental and Applied Physics, University of Regensburg},
            addressline={Universitätsstraße 31}, 
            postcode={93053},
            city={Regensburg}, 
            country={Germany}}

\begin{abstract}
Semiconducting transition metal dichalcogenides (TMDCs) present new possibilities for designing novel electronic devices. An efficient contacting scheme is required to take advantage of exceptional opto-electronic properties of TMDCs in future electronic devices. This is however challenging for TMDCs, mostly due to the typically high Schottky barrier formed between a metal and a semiconductor. Here we investigate different approaches for contacting MoS$_{2}$, utilizing both metallic gold and semimetallic bismuth as contact materials. The collected I-V characteristics of Bi-contacted devices are compared with the performance of traditional gold contacts. The method of AFM ironing, which we used to enhance the parameters of gold contacts, is also described. Additionally, we show preliminary results regarding an optical response for both types of samples.
\end{abstract}




\end{frontmatter}



\section{Introduction}
\label{introduction}

The past decade has seen a growing interest in semiconducting van der Waals (vdW) materials because of their unique electronical and optical properties, which make them attractive and promising candidates for electronic and opto-electrical devices \cite{Liu2019,Avsar2020,Zollner2019}. One of the intensively investigated groups of two-dimensional (2D) materials is the group of transition metal dichalcogenides (TMDCs), in particular MoS$_{2}$, MoSe$_{2}$, WS$_{2}$ and WSe$_{2}$ \cite{WangBook,KolobovBook,WeeBook}. To fully exploit the potential of these materials by means of electrical transport techniques, an efficient contacting scheme is required, which has proven to be challenging in case of TMDCs \cite{schulman2018,allain2015,zheng2021}. Between a metal and a semiconductor (SC) the Schottky barrier (SB) is typically formed, which is theoretically dependent on the difference between the work function of the metal and the electron affinity of the SC \cite{allain2015}. However, this picture has to be modified in the presence of any states in the bandgap of the SC. When a SC and a metal are in close proximity, the metal's extended wavefunction penetrates into the SC. This means that the SC’s original wavefunctions are hybridised with the wavefunctions of the metal and metal-induced gap states (MIGS) are created in the bandgap. As a consequence, the Fermi level at the metal/SC interface is typically pinned below the edge of the SC's conduction band, which means that the SB is only slightly dependent on the metal used, and the current passing through the interface encounters a significant resistance \cite{schulman2018,zheng2021}. To limit the wavefunction spreading from the metal into the TMDC one can introduce a tunnel barrier in the form of an additional interlayer, e.g., hexagonal boron nitride layer, between the metal and the TMDC \cite{farmanbar2015}. However, any tunnel barrier suppresses significantly the current flow through the junction, as the tunneling current decays exponentially with the barrier thickness. Recent report by Shen et al. \cite{shen2021} presents the possibility of obtaining ohmic contacts between semimetallic bismuth and a semiconducting TMDC layer. Due to the near-zero density of states at the Fermi level of the semimetal, the gap states are sufficiently suppressed, allowing to obtain low-resistance ohmic contacts to TMDC. Additionally to a properly chosen contact material, the quality of the interface is a very important factor in realizing good electrical contacts. Transfer techniques commonly used to stack individual 2D layers into vdW structures can lead to inhomogeneities due to contaminants trapped between a 2D layer and a substrate or a metal layer \cite{haigh2012}, which can further lower a quality of the contact. One of the recently suggested methods to circumvent this problem is the so-called AFM ironing, when a contact-mode AFM scan is used to remove any possible imperfections like organic residues, bubbles or wrinkles, improving adjacency of layers and, therefore lowers the contact’s resistance. This method has been used with success in many experiments \cite{rosenberger2018,kim2019,chen2021}.

In this work we study two different approaches for contacting MoS$_{2}$ monolayer.  For that we prepared a series of samples with gold and bismuth as the contacts' materials (4 samples with gold contacts and 4 samples with bismuth contacts), and tested their electrical performance. In order to check how both types of contacting schemes affect emission properties of a MoS$_{2}$ layer, we also performed photoluminescence measurements.

\section{Results and discussion}
\subsection{Investigated samples}

\begin{figure}[H]
    \centering
    \includegraphics[width=0.49\textwidth]{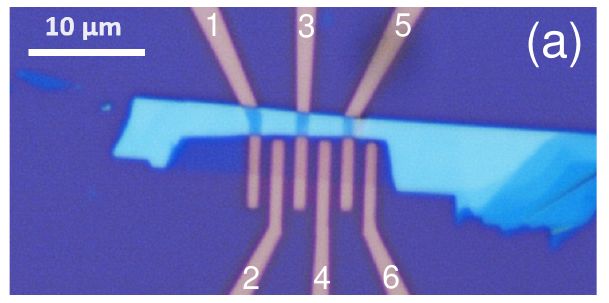}
    \includegraphics[width=0.49\textwidth]{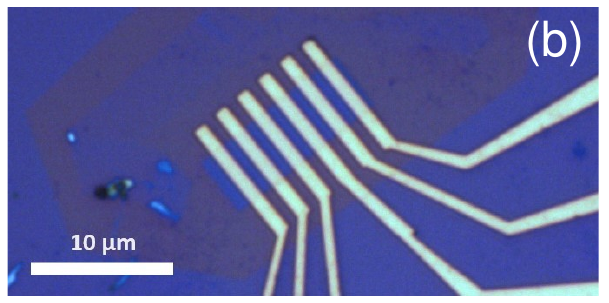}
    \caption{(a) Optical image of sample A with gold contacts. All contacts are 1 $\mu$m wide and are 25 nm high, with 1 $\mu$m spaces between them. MoS$_{2}$ monolayer is placed on top of the contacts. Each contact is given a number from range 1 to 6. (b): Optical image of sample B with bismuth contacts. All contacts are 1 $\mu$m wide and are 100 nm high, with 1 $\mu$m spaces between them. MoS$_{2}$ monolayer is placed below the contacts. }
    \label{fig1}
\end{figure}

Figure \ref{fig1} shows two of the investigated samples. The panel (a) shows the sample A -- a back-gated MoS$_{2}$ monolayer with gold contacts. MoS$_{2}$ flakes were produced by means of mechanical exfoliation from bulk crystals (HQ graphene) and transferred onto polydimethylsiloxane (PDMS) gel-films. Then, an optical microscope was used to find monolayers by the optical contrast and the flake height was measured by the atomic force microscope (AFM). Silicon substrates with 285 nm SiO$_{2}$ layer were first cleaned in O$_{2}$ plasma for 5 minutes. Afterwards, the substrates were spin-coated twice with PMMA layers (200k 9\% and 950k 5\%) and baked for 3 minutes in 150 degrees after each process. Subsequently, the contacts pattern was written on each substrate in an electron-litoghaphy process in a Zeiss Auriga SEM. The fine pattern of the contacts was written with 400 $\mu$C/cm$^{2}$ area dose and 20 $\mu$m aperture, and the coarse pattern was written with 350 $\mu$C/cm$^{2}$ area dose and 120 $\mu$m aperture. The pattern was then developed in MIBK:IPA=1:3 solvent and cleaned in isopropanol. Afterwards, so prepared substrates were put in an UHV system where 5 nanometers of titanium and 20 nanometers of gold were evaporated. The PMMA resist was finally resolved in hot acetone (60 degrees). The layer was placed on previously evaporated titanium/gold contacts by the dry transfer method (substrates were heated to 80 degrees). Parts of the flake laying on the contacts were then flattened using the AFM (AFM ironing) to improve adhesion between the layer and the contacts. MoS$_{2}$ flake was one-way-scanned by Park AFM in a contact mode with a non-contact cantilever. The ironing force was set to 100 nN, the scan rate was set to 0.6 Hz , which means the tip velocity was $\sim5~\mu$m/s, and the resolution was set to 512 px, which gives $\sim1$ nm between each line scan. One similar sample was produced, as well as two reference samples with non-ironed contacts. The panel (b) of Fig. \ref{fig1} presents the sample B -- also a MoS$_{2}$ monolayer, with bismuth/gold contacts evaporated on the top of the flake. The monolayer was identified the same way as previously described for the sample A. The layer was then put on the silicon substrate with SiO$_{2}$ backgate, etched with CHF$_{3}$/O$_{2}$ reactive ion etching recipe to obtain a rectangle shape with dimensions: 4.5x12 $\mu$m and finally 20 nm of bismuth and 80 nm of gold were evaporated on the flake in an UHV system, using an electron gun following the same procedure as for the sample A.

\subsection{Electrical measurements}
\begin{figure*}[!ht]
    \centering
    \includegraphics[width=\textwidth]{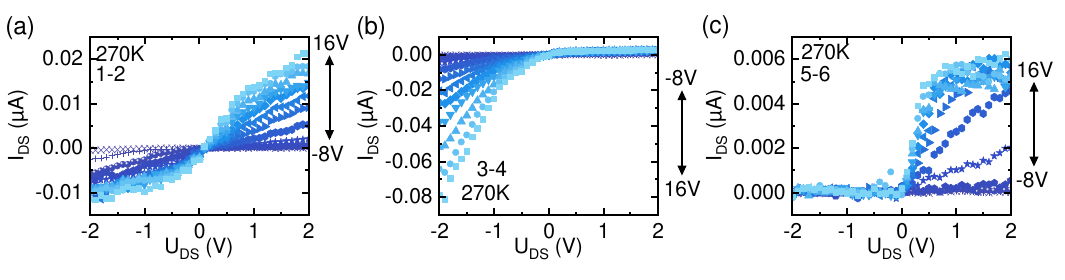}
    \includegraphics[width=\textwidth]{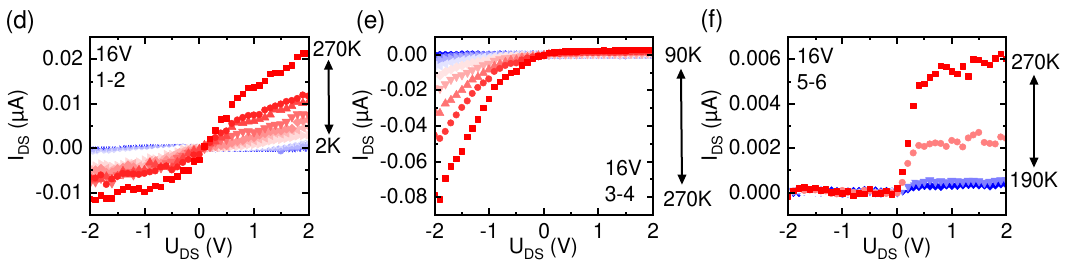}
    \caption{(a)-(c): I-V curves for sample A for backgate voltages in range -8 to 16 V with 2 V step for 270 K. Each graph denotes for which contact pair the I-Vs were measured. Varying curve shapes are visible for different contacts on the same sample, the curves are also asymmetric. The curves' shapes suggest Schottky-type contacts which are also sensitive to the backgate voltage. (d)-(f): I-V curves for sample A for the backgate voltage equal to 16 V in temperature series with 20 K step. Each graph denotes for which contact pair the I-Vs were measured. In case of contacts 1-2 the current was still flowing in 2 K, contacts 3-4 stopped conducting at 90 K and contacts 5-6 stopped conducting at 190 K.}
    \label{fig2}
\end{figure*}
First let us discuss the electrical measurements on the gold-contacted samples. All contacts, six on each sample, were working for both ironed samples. In case of the non-ironed samples only two contacts were working for each sample, which suggests that AFM ironing improved significantly the contacts' performance. The current-voltage (I-V) curves of the sample A for the backgate voltages in range of -10V to 16V and temperatures in range of 2--270 K were measured for each contacts pair (1--2, 3--4 and 5--6). The results are presented in Fig. \ref{fig2}. Analysing the I-V curves shape, we note that all contacts have Schottky-like characteristics and the curves are changing with the backgate voltage. We can also observe that the curves vary for each contact pair and are asymmetric for negative and positive drain-source voltages, which suggests different barriers for different contacts. With decreasing temperature the drain-source current also decreases, which is typical for Schottky contacts, as the current density is described by thermal transport model \cite{Somvanshi2017}: 
\begin{equation}
    J = J_{S}\exp\left(\frac{-q\beta U}{k_{B}T}\right)\left[\exp\left(\frac{qU}{k_{B}T}\right)-1\right],
\end{equation}
where $J_{S}=AT^{3/2}$ is the saturation current density, $A$ is the Richardson constant, $T$ is the temperature, $q$ is the elemental charge, $\beta=1-\frac{1}{n}$, $n$ is the contact ideality parameter (equals 1 for an ideal Schottky contact), $U$ is the applied voltage and $k_{B}$ is the Boltzmann constant. In our case, only one contact pair was still conducting at 2 K (Fig. \ref{fig2}(d)), the second contact pair stopped conducting at 90 K and the third contact pair stopped conducting at 190 K. This behaviour is in agreement with our expectations as the thermal transport is substantially decreased at lower temperatures. We analysed I-Vs in a wide range of temperatures to calculate SB heights for each contact by using a thermal transport model (eq. 1) for two back-to-back Schottky diodes. The SB height was derived from the formula below \cite{Somvanshi2017}:
\begin{equation}
    \phi_{B}=\phi_{B}^{eff}-U\left(1-\frac{1}{n}\right)=-\frac{k_{B}}{q}\frac{d \ln \left(\frac{J}{T^{3/2}}\right)}{d\frac{1}{T}}-U\left(1-\frac{1}{n}\right),
\end{equation}
where $\phi_{B}^{eff}$ is the effective SB -- the barrier measured with a drain-source voltage applied. 
\begin{figure*}[ht]
    \centering
    \includegraphics[width=0.65\textwidth]{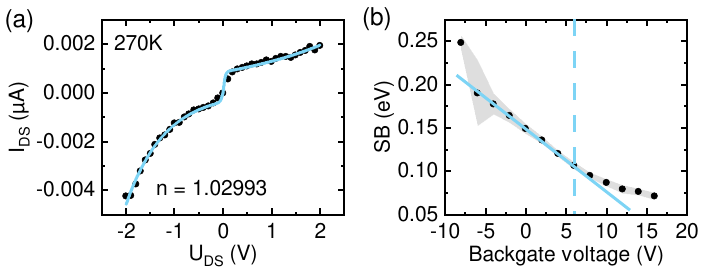}
    \caption{(a): I-V curve of contacts 3-4 for zero backgate voltage at 270 K (black dots) with fitted formula from eq. 3 (blue line). Ideality parameter is extracted from fit of the I-V part for negative drain-source voltage. (b): SB height as a function of applied backgate. True SB height is the value for 6V backgate.}
    \label{fig3}
\end{figure*}
The barrier and the ideality parameter $n$ was extracted from fitting the I-V curve for zero backgate voltage with the formula below:
\begin{equation}
    J=\frac{J_{S}\sinh\left(\frac{qU}{2k_{B}T}\right)}{\cosh \left(\frac{q(1-\frac{2}{n})U}{2k_{B}T}\right)}
\end{equation}

The obtained ideality parameter equals 1.02993(61) for the part of the I-V for negative drain-source voltages, as this part was used to calculate the SB height (Fig. \ref{fig3}(a)). It was possible to make calculations for one contact only, because for all other contacts the drain-source current was very low, which led to huge uncertainties of the derived values. We fitted linearly the Arrhenius plot of $\ln(I/T^{3/2})$ for the drain-source voltage equal to -1.9 V and calculated the effective SB heights for every backgate voltage using eq. 2. The true SB was determined for the backgate voltage equal 6 V, as for this voltage we obtain the flat-band condition (Fig. \ref{fig3}(b)) \cite{Somvanshi2017}. The final SB height is equal 0.1617(41) eV with correction of $n$ parameter (calculated using the eq. 2).

The average SB height reported in the literature for Au/MoS$_{2}$ equals $\sim$0.19 eV \cite{allain2015}. This means we were able to obtain the SB height slightly lower than the average value reported, which we believe is due to the AFM ironing performed on the contacts. However, the overall results are not very satisfying as the current passing through the contacts is very low even at room temperature.

\begin{figure*}[!htbp]
    \centering
    \includegraphics[width=\textwidth]{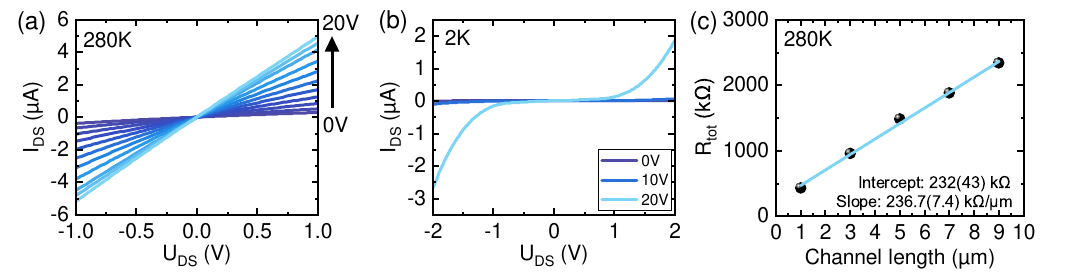}
    \caption{(a): I-V curve for sample B for backgate voltages from 0 to 20 V at 280 K. The contacts have ohmic behaviour. (b): I-V curve for sample B for 0, 10 and 20 V backgate voltages at 2 K. The ohmic behaviour is lost. (c): Measured total resistance as a function of a channel length at 280 K for a zero backgate voltage. The resistance increases linearly with increasing the distance between the contacts.}
    \label{fig4}
\end{figure*}

Let us now discuss the results of the measurements on the sample B. We measured I-V curves for temperatures from 2 K to 280 K and for different backgate voltages. Fig. \ref{fig4}(a) presents the I-V characteristics at 280 K for backgates in range 0 V -- 20 V. The curves show ohmic behaviour of Bi/MoS$_{2}$ contacts at room temperature with resistance decreasing with the increasing backgate voltage. The current passing through this sample is substantially higher than through the sample A. The contacts lose their ohmic characteristics at temperatures below 100 K. Such a behaviour of I-V curves was typical for all our devices with bismuth contacts. The I-V curves for 0 V, 10 V and 20 V backgate voltages at 2 K are presented in Fig. \ref{fig4}(b)). We can observe much smaller current passing through the sample. This can be explained by the carriers freezing in cryogenic temperatures and a high backgate voltage is needed to see a current flowing. We plotted the total resistance as a function of the channel length at 280 K. One can observe the linear dependence of the resistance on the channel length, as expected assuming the uniform sheet resistance of MoS$_{2}$. We calculated the contact and channel resistances in 280 K using transfer length method \cite{Scott1982}. The total resistance of the investigated sample is described by the formula:
\begin{equation}
    R_{tot}=2R_{c}+R_{ch}=2R_{c}+R_{sh}L_{ch},
\end{equation}
where $R_{c}$ is the contact resistance, $R_{ch}$ is the channel resistance, $R_{sh}$ is the sheet resistance of the MoS$_{2}$ channel and $L_{ch}$ is the channel length. From the intercept of the linear fit one can extract the resistance for "zero" channel length which equals the resistance of two contacts. In case of the sample B we obtain $R_{c}=522(99)$ k$\Omega\cdot\mu$m. This is a very high value for a MoS$_{2}$ contact \cite{Kim2024}, being three orders of magnitude higher than the best value obtained for a bismuth/MoS$_{2}$ contact \cite{shen2021}. The possible reasons for this might be non-ideal evaporation parameters or the residues of PMMA resist used during electron litography processes. The sheet resistance calculated from the slope equals $R_{s}=1065(34)$ k$\Omega/\square$. The obtained value is similar to the result obtained by Yang et al. \cite{Yang2018}. However, it is two orders of magnitude higher than the values usually reported for a monolayer MoS$_{2}$ \cite{Zhan2012,shen2021}. This might suggests much smaller electron mobility or carrier density than usually measured in MoS$_{2}$.

\subsection{Optical measurements}

In order to further compare the performance of both conducting schemes we also studied how they affect the optical properties of the MoS$_{2}$ layer. For that purpose the samples were investigated by measuring microphotoluminescence spectra at cryogenic temperatures. Measurements were performed in an optical helium flow cryostat (Oxford Instruments). The excitation source was a 532 nm continuous-wave laser. The laser light was focused on the sample by 0.75 NA, and 50× microscope objective (Mitutoyo). The output signal was then directed to the spectrometer (500 mm focal length, 300 groove/mm grating) equipped with a liquid-nitrogen-cooled CCD camera (Princeton Instruments). Fig. 5(a) presents a representative PL spectrum of one of the Au-contacted samples at 4.5 K, whereas Figure 5(b) shows a representative PL spectrum of one of the Bi-contacted samples at the same temperature. Both spectra are compared to  a PL spectrum of a bare MoS$_{2}$ flake transferred on SiO$_{2}$, taken with the same laser power of 0.1 mW. On both spectra we can observe typical emission lines for a MoS$_{2}$ monolayer - negative trion A - T$\mathrm{_{A}}$, neutral exciton A - X$\mathrm{_{A}}$ and defect-related emission - D \cite{Jadczak2017}. The lines were identified by comparing the measured spectra with the spectrum of the reference flake and by analysing the reflectivity from both samples. We can also observe substantial broadening of the emission lines suggesting that the emission is very inhomogeneous. Particularly a line shape observed for a Bi-contacted sample is quite unusual for a MoS$_{2}$ monolayer, as it suggests that there is an additional emission peak L at $\sim1.86$ eV. We note that it is observed only on the Bi-containing samples thus we hypothesize it is connected with the process of bismuth evaporation. The detailed origin of this new transition is a subject of ongoing investigations.

\begin{figure}[H]
    \centering
    \includegraphics[width=\textwidth]{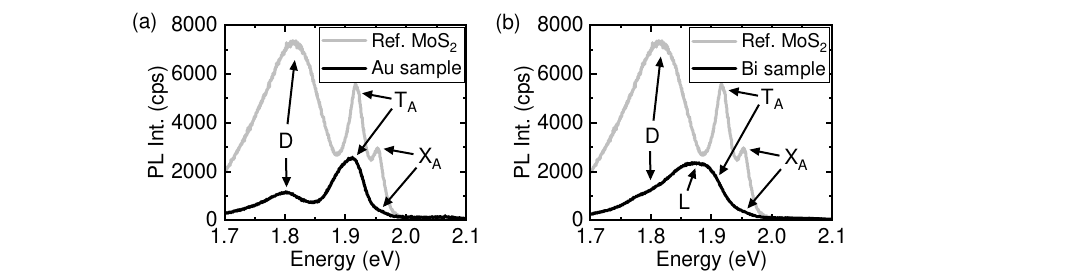}
    \caption{(a): Representative  PL spectrum of a sample with gold contacts at 4.5 K compared with a PL spectrum of a bare MoS$_{2}$ flake. All typical emission lines for a MoS$_{2}$ monolayer (T$\mathrm{_{A}}$, X$\mathrm{_{A}}$ and D) are visible for the sample with Au contacts, although the lines are broader and the overall PL intensity is smaller. (b): Representative PL spectrum of a sample with bismuth contacts at 4.5 K compared with a PL spectrum of a bare MoS$_{2}$ flake. The spectrum shape is very unusual for a MoS$_{2}$ monolayer and indicates the presence of an additional emission peak L.}
    \label{fig5}
\end{figure}

\section{Conclusions}
We discussed two approaches for contacting MoS$_{2}$ monolayers. In case of metal gold contacts we obtained clear Schottky I-V characteristics and extracted the SB height being equal to 0.1617(41) eV, what makes it slightly lower than the average literature value. In case of semimetal bismuth contacts, we obtained ohmic behaviour in temperature range from 280 K down to 100 K, however, with quite high contact resistance. We can confirm that bismuth is better than gold for contacting MoS$_{2}$, although this contacting scheme requires a deeper investigation to obtain ohmic behaviour also at cryogenic temperatures. The influence of Bi contacts on the optical properties of the MoS$_{2}$ monolayers also need to be studied further as the PL spectrum of the corresponding samples shows unusual features, which could provide us with more information on Bi/MoS$_{2}$ interfaces.

\section*{Acknowledgements}

The studies were financially supported from the grant no. 2022/45/B/ST5/04292 funded by The National Science Centre in Poland. We also acknowledge financial support from European Union (EU) student exchange programme Erasmus+ Internship.












\end{document}